# A model for a Lindenmayer reconstruction algorithm

Diego Gabriel Krivochen & Beth Phillips

Abstract:

Given an input string *s* and a specific Lindenmayer system (the so-called *Fibonacci grammar*), we define an automaton which is capable of (i) determining whether *s* belongs to the set of strings that the Fibonacci grammar can generate (in other words, if *s* corresponds to a generation of the grammar) and, if so, (ii) reconstructing the previous generation.

We assume a two-tape one-way finite automaton (2FA):

**Definition 1:** a two-tape one-way deterministic FA (2FA) is a tuple $\langle \Sigma, Q, \delta, q_0, F \rangle$, where:

- $\Sigma$ is the input alphabet. $\Sigma = \{0, 1\}$
- $Q$ is the set of states
- $q_0 \in Q$ is the initial state
- $\delta$ is a transition function
- $F \subseteq Q$ is the set of accepting states
- $\triangleright, \triangleleft \notin \Sigma$ are left- and right-end markers, respectively

Each tape has an independent head. At each point, a set of rules determines the action and movement of the heads depending on what is written on the input tape (Tape 1) and determines what is to be written in the output tape (Tape 2), which is initially empty (which we will indicate with the symbol $\varepsilon$).

**Definition 2:** given a 2FA $G$, the *configuration* of $G$ is a tuple $\langle x_1, x_2, q_n, i_1, i_2 \rangle \in (\triangleright\Sigma^*\triangleleft)^2 \times Q \times \mathbb{N}^2$, where $q_n \in Q$ is the current state, $x_1$ and $x_2$ are the contents of tapes 1 and 2 respectively and $i_1$ and $i_2$ are the positions of the 1st and 2nd heads[1]. We can simplify this by assuming that $x_n = i_n$: because the only symbol that each head is 'aware' of is the currently scanned one, specifying the content of the tape implies that the head is at that position.

Because our 2FA is *one way*, the heads can do two things as far as movement is concerned:

R: move one slot to the right
S: stay in place

There is no leftward movement.

The transition function for our 2FA is a rule of the form

$$\delta(q_n, x_1, x_2) = (q_m, y_1, y_2, M_1, M_2)$$

Note that the rule specifies the input state, the contents of T1 and T2, and the machine's movements (denoted by 'M') in each tape; as we said above, there is no need to specify the position of the head separately. Specifically, the transition function in generalized form can be read as 'in state $q_n$, with $x_1$ in T1 and $x_2$ in T2, proceed to state $q_m$ by changing $x_1$ to $y_1$ and $x_2$ to $y_2$, then move or stay in place on T1 and move or stay in place on T2'.

For example,

$$\delta(q_1, 0, \varepsilon) = (q_2, 0, 0, R, S)$$

---

[1] Furia (2012) offers a generalized definition which, for our present purposes, seems to be an overkill.

means 'in state $q_1$, with 0 on T1 and $\varepsilon$ on T2, proceed to state $q_2$; leave 0 on T1 (or, equivalently, replace 0 by 0 on T1) and move right, and replace $\varepsilon$ by 0 in T2 and stay in place'.

Tape 1 contains a string of 1s and 0s, Tape 2 is initially empty. The head in T1 is read-only (or, equivalently, it rewrites every symbol as itself), the head in T2 is read-write. It is crucial to note that the content of Tape 2 at $i_n$ will depend on the content of T1 at $i_n$ as per the corresponding rule.

**Definition 3:** a *multitape automaton* with *n*-tapes A is *k*-synchronized if for *n* heads $h_1, \ldots, h_n$, these heads are never farther than *k* apart (Ibarra & Tran, 2012).

The heads in our 2FA are *2-synchronized*. This limit is given by the size of the minimal grammatical constituents in the strings evaluated by the automaton: the minimal constituent size we will consider is 1 and the maximal constituent size is 2, as defined by the rules of the grammar which generate the input string. The very specific constraints set on our 2FA derive from the specificity of what we want it to do.

The 2FA we construct here does two things:

- It serves as a test for Fibonacci-membership applied to the output of a generative Lindenmayer system with alphabet $\Sigma = \{0, 1\}$ and rules $0 \rightarrow 1; 1 \rightarrow 0\ 1$ (cf. Prusinkiewicz & Lindenmayer, 2010) –*Fib grammar* henceforth- (operations over Tape 1)

A derivation for the Fib grammar is illustrated as follows:

Fibonacci sequence = {0, 1, 1, 2, 3, 5, 8, 13, 21…}

$$0 \rightarrow 1 \text{ symbol}$$
$$1 \rightarrow 1 \text{ symbol}$$
$$01 \rightarrow 2 \text{ symbols}$$
$$101 \rightarrow 3 \text{ symbols}$$
$$01101 \rightarrow 5 \text{ symbols}$$
$$10101101 \rightarrow 8 \text{ symbols}$$
$$0110110101101 \rightarrow 13 \text{ symbols}$$
$$101011010110110101101 \rightarrow 21 \text{ symbols}$$

…

The choice of this particular Lindenmayer system to generate the Fibonacci sequence finds justification in how low-level local transition facts map to higher-level constituent structure: the fact that a [0] is always followed by a [1] allows us to 'group' [01] units when reading an output string from the Fib grammar from left-to-right and map them to their previous generation, which as per the rules of the grammar is [1]. Relevantly, this reconstruction property does not extend to the other irreducible Fibonacci Lindenmayer grammar (which we refer to as *bif*), $0 \rightarrow 1; 1 \rightarrow 1\ 0$: since [1] in a generation $g_n$ can be mapped to a [0] in $g_{n-1}$ by itself, a point of ambiguity is created at every [1] in the string; furthermore local bi-grams [01] do not longer correspond to a constituent. Low-level and high-level properties of the grammar (or *representational* and *derivational* properties) do not map to each other in the same way for the Fib and *bif* grammars.

Applied only once to a string *s*, the 2FA proposed here assess only the 'Fibonacci-grammaticality' of *s*. By 'Fibonacci-grammaticality' we mean 'compliance with the following local *n*-gram constraints':

*111
*00

It must be noted that a string may be Fibonacci-grammatical *without* it being an actual Fibonacci string (i.e., an output of the Fib grammar). The 2FA does this by having the transitions corresponding to these *n*-grams undefined, with which the procedure terminates immediately.

**Theorem:** a string *s* corresponds to a generation of the *Fibonacci grammar* if and only if after a finite number of recursive applications of the transition functions 1-3 specified below, the algorithm stops when there is only a 0 (the axiom of the Fibonacci grammar) on the output tape.

This means that 2FA can in fact determine the Fibonacci membership of a string as long as the length of that string (notated |*s*|) is a Fibonacci number.

- Given any generation $g_n$ of the *Fib grammar*, 2FA can construct $g_{n-1}$ on Tape 2

**Instructions:**

1. $\delta(q_1, 0, \varepsilon) = (q_2, 0, 0, R, S)$

In state $q_1$, with 0 on T1 and $\varepsilon$ on T2, proceed to state $q_2$; leave 0 in T1 (or, equivalently, replace 0 by 0 on T1) and move right, and replace $\varepsilon$ by 0 in T2 and stay in place.

2. $\delta(q_3, 1, 0) = (q_4, 1, 1, R, R)$

In state $q_3$, with 1 on T1 and 0 on T2, proceed to state $q_4$; leave 1 on T1 (or, equivalently, replace 1 by 1 on T1) and move right, and replace 0 by 1 on T2 and move right.

3. $\delta(q_5, 1, \varepsilon) = (q_6, 1, 0, R, R)$

In state $q_5$, with 1 on T1 and $\varepsilon$ on T2, proceed to state $q_6$; leave 1 on T1 (or, equivalently, replace 1 by 1 on T1) and move right, and replace $\varepsilon$ by 0 on T2 and move right.

Recall that T1 is (presumably) a Fib string; T2 is initially ▷ $\varepsilon$ $\varepsilon$ $\varepsilon$ …◁.

The algorithm operates until $h_1$ in T1 reaches ◁ in an accepting state. When this happens, $h_2$ in T2 moves right until reaching ◁. Because 2FA is *2-synchronized*, if $h_1$ reaches ◁ in *n* steps, $h_2$ will reach ◁ in at most *n+2* steps. An input string *s* is accepted if both heads end up on ◁ after a finite number of steps.

**Example:**

We will mark in **bold** the symbols in T1 that have already been read at each point. In this example, T1 contains generation 4 of the *Fib-grammar*.

1) T1 ▷**0**1101◁
   T2 ▷0 $\varepsilon$ $\varepsilon$ $\varepsilon$ $\varepsilon$ ◁ (by 1)

2) T1 ▷**01**101◁
   T2 ▷1 $\varepsilon$ $\varepsilon$ $\varepsilon$ $\varepsilon$ ◁ (by 2)

3) T1 ▷**011**01◁
   T2 ▷10 $\varepsilon$ $\varepsilon$ $\varepsilon$ ◁ (by 3)

4) T1 ▷**0110**1◁
   T2 ▷100 $\varepsilon$ $\varepsilon$ ◁ (by 1)

5) T1 ▷**01101**◁
   T2 ▷101 $\varepsilon$ ◁ (by 2)